\documentclass[twocolumn]{openjournal}
\usepackage{natbib}
\usepackage{graphicx,amsmath,amssymb,amstext}
\usepackage{amsbsy,amsfonts,amsthm,color}
\usepackage[colorlinks,linkcolor=blue,citecolor=blue,urlcolor=blue ]{hyperref}
\usepackage[utf8]{inputenc}
\usepackage{float}
\usepackage[caption=false]{subfig}

\defcitealias{Prole2023}{LP23}
\defcitealias{Schauer2021}{AS21}

\begin{document}


\title{Population III star formation: multiple gas phases prevent the use of an equation of state at high densities.\vspace{-3em}}

\author{Lewis R. Prole$^{* \ 1,2}$}
\author{Paul C. Clark$^{2}$}
\author{Felix D. Priestley$^{2}$}
\author{Simon C. O. Glover$^{3}$}
\author{John A. Regan$^{1}$}
\email{$^*$email: lewis.prole@mu.ie}

\affiliation{Centre for Astrophysics and Space Sciences Maynooth, Department of Theoretical Physics, Maynooth University, Maynooth, Ireland.}
\affiliation {Cardiff University School of Physics and Astronomy}
\affiliation {Universit\"{a}t Heidelberg, Zentrum f\"{u}r Astronomie, Institut f\"{u}r Theoretische Astrophysik, Albert-Ueberle-Stra{\ss}e 2, D-69120 Heidelberg, Germany.}


\begin{abstract}

\noindent Advanced primordial chemistry networks have been developed to model the collapse of metal-free baryonic gas within the gravitational well of dark matter (DM) halos and its subsequent collapse into Population III stars. At the low densities of 10$^{-26}$-10$^{-21}$ g cm$^{-3}$ (10$^{-3}$-10$^2$ cm$^{-3}$) the collapse is dependent on H$_2$ production, which is a function of the compressional heating provided by the DM potential. Once the gas decouples from the DM, the temperature-density relationship follows a well established path dictated by various chemical reactions until the formation of the protostar at 10$^{-4}$ g cm$^{-3}$ (10$^{19}$ cm$^{-3}$). Here we explore the feasibility of replacing the chemical network (CN) with a barotropic equation of state (EoS) just before the formation of the first protostar, to reduce the computational load of simulating the further fragmentation, evolution and characteristics of the very high density gas. We find a significant reduction in fragmentation when using the EoS. The EoS method produces a protostellar mass distribution that peaks at higher masses when compared to CN runs. The change in fragmentation behaviour is due to a lack of cold gas falling in through the disc around the first protostar when using an EoS. Despite this, the total mass accreted across all sinks was invariant to the switch to an EoS, hence the star formation rate (M$_\odot$ yr$^{-1}$) is accurately predicted using an EoS. The EoS routine is approximately 4000 times faster than the CN, however this numerical gain is offset by the lack of accuracy in modelling secondary protostar formation and hence its use must be considered carefully.

\end{abstract}


\maketitle

\section{Introduction} 
\label{sec:intro}


\noindent Small-scale simulations investigating Pop III star formation in individual star-forming halos have made considerable progress over the last decade. However, they remain unable to follow the build-up of the Pop III IMF over a long enough period to provide definitive results. Models with resolutions high enough to resolve individual protostars ($\Delta$x$_{\rm cell}$ $\sim$ 0.1-0.01 au) can be run for only short periods (100 - 1000 yr; see e.g. \citealt{Greif2012,Prole2022,Prole2022a,Hirano2022,Prole2023}). Lower resolution models ($\Delta$x$_{\rm cell} \gtrsim$ 1 au) can be run for longer periods, often 10$^3$ - 10$^4$ years or more \citep[see e.g.][]{Stacy2013,Susa2014,Stacy2016,Wollenberg2020,Sharda2020,Jaura2022}, but these studies underestimate fragmentation within the gas and hence overestimate the resulting stellar masses, owing to the absence of fragmentation-induced starvation \citep{Peters2010,Machida2013,Prole2022}.

A large obstacle to simulating Pop III star formation are the vast computational resources required. This is largely due to the absence of the so-called first Larson core \citep{Larson1969}. In present-day star formation, this core forms at a density of $10^{-13} \: {\rm g \: cm^{-3}}$ once the gas becomes optically thick to its own dust emission, leading to the further collapse of the gas becoming adiabatic and hence stable to fragmentation \citep[e.g.][]{Masunaga1998}. In metal-free gas this transition to adiabatic evolution does not occur until it reaches a density of $\sim 10^{-4} \: {\rm g \: cm^{-3}}$, almost 10 orders of magnitude higher \citep{Omukai2000}. Pop III gas therefore remains susceptible to fragmentation over a much broader range of densities, all of which must be resolved in order to obtain converged results. \\
\indent In order to resolve these high densities, small cell sizes are required owing to the Truelove criterion -- the requirement that the Jeans length be resolved with at least four cells in order to avoid artificial fragmentation \citep{Truelove1997}. This in turn implies a very short timestep, owing to the limitation set by the Courant condition \citep{Courant1952}, i.e. the timestep must be sufficiently short such that information can not travel at the sound speed across the whole cell during the timestep, ensuring that information from a cell can only be communicated to its immediate neighbours. 

For a mesh cell with a size equal to the Jeans length $\lambda_J$, the largest possible stable timestep is
\begin{equation}
\delta t = \frac{\lambda_J}{c_{\rm s}} \sim \sqrt{ \frac{1}{G \rho }},
\end{equation}
where $c_{\rm s}, \rho$ and $G$ are the sound speed, density and gravitational constant, respectively. For cells close to the protostellar density of 10$^{-6}$ g cm$^{-3}$ this corresponds to $\delta t\sim0.1$ yr. Furthermore, most simulations resolve $\lambda_J$ by at least 16 cells, reducing  $\delta t$ by an additional order of magnitude.

There is therefore a need to reduce the computing time per timestep while simulating Pop III star formation, in order to properly characterise the mass of the first stars accurately over reasonable timescales. One aspect of the computation that is becoming increasingly more expensive is the chemical network needed to track the heating and cooling of the gas as it is violently pulled into the gravitational potential of the dark matter (DM) halo. Here the gas is shock heated up to $\sim$1000 K where it can produce the coolant H$_2$, which allows it to collapse and decouple from the dark matter. From there, the collapse is complicated by many reactions. While H$_2$ is initially formed via the slow radiative association of H atoms and electrons producing H$^-$ followed by a fast associative detachment reaction to form H$_2$, the formation rate is dependent on the ionization fraction and hence decreases as the gas recombines \citep{Glover2006}. This allows three-body reactions to take over as the primary source of H$_2$ production \citep{Palla1983}. These rapidly convert all of the hydrogen to H$_{2}$ once the gas density exceeds $\rho \sim 10^{-15} \: {\rm g \, cm^{-3}}$. However, following an initial boost to the cooling rate associated with this chemical transition, H$_{2}$ line cooling becomes increasingly inefficient as the density increases, owing to the growing optical depth of the H$_2$ rovibrational lines \citep{Ripamonti2004,Turk2011}. At a density of $\rho \sim 10^{-10} \: {\rm g \, cm^{-3}}$, collision-induced emission kicks in to become the dominant cooling process until the gas is hot and dense enough to dissociate H$_2$ \citep{Omukai2005,Yoshida2008}, which provides further cooling until it is depleted and the collapse becomes adiabatic at the formation of the protostar at 10$^{-4}$ g cm$^{-3}$.

Given that chemistry arguably plays a more important role in primordial star formation compared to present day star formation \citep{Glover2012}, understanding the chemical reactions that occur between the few chemical species available in primordial gas is crucial to understanding Pop III star formation. For example, \cite{Glover2006} showed that the uncertainties in associative detachment and mutual neutralisation rate coefficients introduce large uncertainties in the H$_2$ abundance and cooling rate of hot ionised gas during the collapse. Likewise, \cite{Turk2011} showed that uncertainties in the three-body formation rate drastically changes the chemical abundances, morphology and velocity structure of the gas in high density regions. Additionally, \cite{Bovino2013} showed that using a low-order primordial chemistry solver leads to a resolution dependence in the radial profiles of various chemical abundances, concluding that accurate modelling of the chemistry and thermodynamics is central for simulating primordial star formation.

Despite these complications, the temperature-density relationship follows a well documented path  throughout the collapse (e.g. \citealt{Omukai2000,Omukai2005,Yoshida2006,Greif2008,Clark2011,Prole2022}), as we will show in \S \ref{sec:baro}. This prompts the question of whether a full chemical network is necessary once the gas has decoupled from the dark matter. As previous studies have opted to use a barotropic EoS   to handle the temperature-density relationship (e.g. \citealt{Saigo2004,Suwa2007a,Machida2008a,Bonnell2008,Machida2015,Riaz2018,Susa2019,Hirano2022,Raghuvanshi2023}), here we aim to investigate what the limitations are of employing an EoS instead of using a full chemical network (CN). In particular, we examine the impact that using an EoS has on subsequent fragmentation compared to the full CN as well as the overall star formation rate. To that end, we re-simulate the star forming halos from \cite{Prole2023} (hereafter \citetalias{Prole2023}) using a barotropic EoS designed to reproduce the same density-temperature relationship found in that study. The phase of the simulations that requires the most computational resources is following the fragmentation behaviour of the gas and sink particle formation/accretion once the highest refinement has been reached, so here we focus on this phase only by switching to a barotropic EoS at a time just before the formation of the first protostar.

The structure of the paper is as follows: In \S \ref{sec:method} we discuss the numerical method, our use of sink particles, the CN simulations of \citetalias{Prole2023}, and the construction of our empirical EoS. In \S \ref{sec:results} we present the differences in fragmentation and accretion behaviour between the CN and EoS methods and discuss the physical causes for the differences. In \S \ref{sec:caveats} we discuss caveats before summarising in \S \ref{sec:summary}.

\ \
\section{Numerical method}
\label{sec:method}

\begin{figure*}
	 \hbox{\hspace{0cm} \includegraphics[scale=0.7]{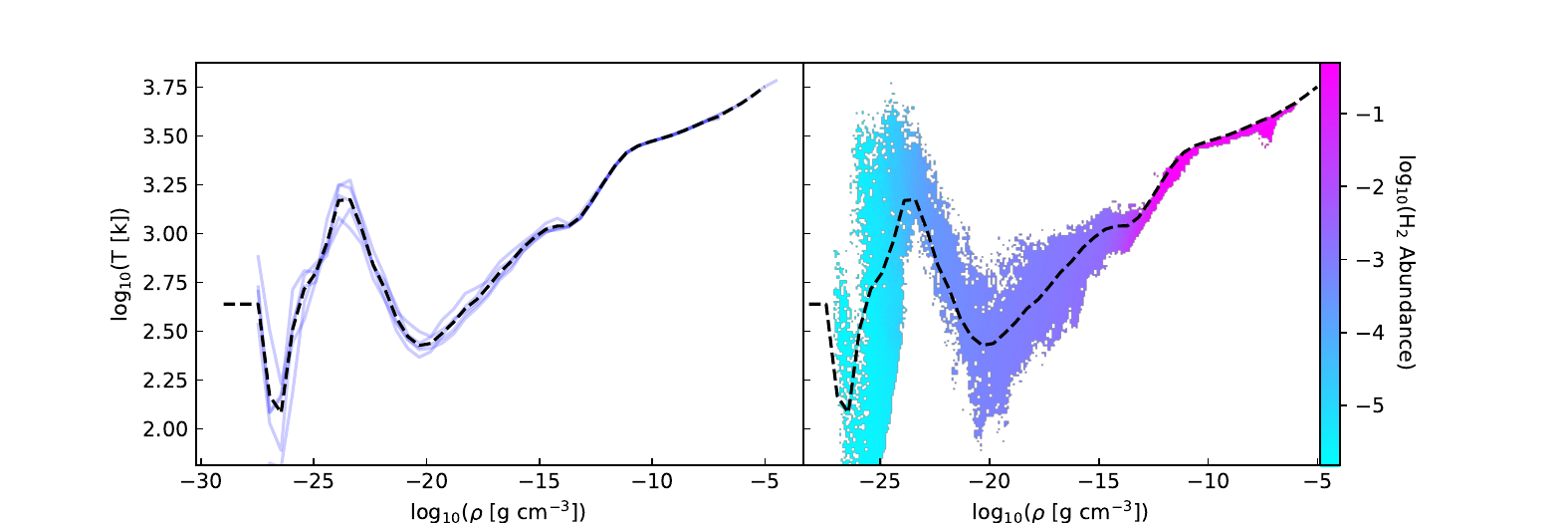}}
    \ \
    \caption{Barotropic EoS used. Left: mass weighted average temperature-density profiles of the 5 cosmological halos (Halos A, B, C, D \& E) of \citetalias{Prole2023} shown in blue, with the average profile overplotted in dashed black. Right: The constructed empirical EoS contrasted against the temperatures-density profile of Halo E when using the chemical network, coloured by H$_2$ mass fraction.}
    \label{fig:eos}
\end{figure*}
\subsection{Cosmological initial conditions}
\noindent To directly compare the use of an EoS versus the chemical network (CN), we re-simulate 5 cosmological halos from \citetalias{Prole2023}, which were selected from the cosmological simulations performed by \cite{Schauer2021} (which we name as Halos A, B, C, D \& E). We chose to only consider the suite that did not include a Lyman-Werner background radiation field, as this allows a more straight forward comparison. These cosmological simulations assumed a $\Lambda$CDM cosmology with parameters $h=0.6774$, $\Omega_0 = 0.3089$, $\Omega_{\rm b} = 0.04864$, $\Omega_\Lambda = 0.6911$, $n = 0.96$ and
$\sigma_8 = 0.8159$ as derived by the \cite{Planck-Collaboration2020}.  The simulations were initialised at z=200 with the initial DM distribution created by MUSIC \citep{Hahn2011} using the transfer functions of \cite{Eisenstein1998} and the gas distribution initially followed the DM. While the DM and gas components were initially made up of 1024$^3$ cells each within the 1 Mpc $\rm{h^{-1} (1+z)^{-1}}$ comoving box, the gas component had an additional continuous refinement criteria of 16 cells per Jeans length until the creation of sink particles above a threshold density of $\sim 10^{-19}$ g cm$^{-3}$. 

\subsection{Zoom-in simulations and sink particles}
\noindent For both the CN and the EoS high resolution zoom-in simulations, we continuously refine the mesh so that the Jeans length is resolved by at least 16 cells down to a minimum cell length of 0.028 AU. Sink particles are inserted into the simulations at a threshold density of 10$^{-6}$ g cm$^{-3}$ to prevent artificial collapse when the simulation reaches its maximum refinement level. Our sink particle implementation was introduced in \cite{Wollenberg2019} and \citet{Tress2020}. A cell is converted into a sink particle if it satisfies three criteria: 1) it reaches a threshold density; 2) it is sufficiently far away from pre-existing sink particles so that their accretion radii do not overlap; 3) the gas occupying the region inside the sink is gravitationally bound and collapsing. Likewise, for the sink particle to accrete mass from surrounding cells it must meet two criteria: 1) the cell lies within the accretion radius; 2) it is gravitationally bound to the sink particle. A sink particle can accrete up to 90$\%$ of a cell's mass, above which the cell is removed and the total cell mass is transferred to the sink.

In both the EoS and the CN runs, we choose the initial sink particle accretion radius $R_{\text{sink}}$ to be the Jeans length $\lambda_{\text{J}}$ corresponding to the sink particle creation density. We take the value from \cite{Prole2022} of $1.67 \times 10^{12}$ cm for the sink accretion radius. We set the minimum cell length to fit 8 cells across the sink particle accretion radius in compliance with the Truelove condition, giving a minimum cell volume $V_{\text{min}}=(R_{\text{sink}}/4)^3$. The minimum gravitational softening length for cells and sink particles, $L_{\text{soft}}$, is set to $R_{\text{sink}}/4$. 

We also include the treatment of sink particle mergers used in \citet{Prole2022}. Briefly, we allow sinks to merge if they fulfil four criteria: 1) they lie within each other's accretion radius;  2) they are moving towards each other; 3) their relative accelerations are $<0$; and 4) they are gravitationally bound to each other.

As in \citetalias{Prole2023}, the sink particle accretion radius grows in time with on-the-fly calculations of the stellar radius using an approximate analytic formulae taken from \cite{Hosokawa2012}\footnote{Equation was originally derived and presented by \cite{Stahler1986} but with a typo mistaking the normalisation of the accretion rate as 10$^3$ M$_\odot$ yr$^{-1}$ instead of  10$^{-3}$ M$_\odot$ yr$^{-1}$}
\begin{equation}
\rm{R_{\text{sink}}=26 R_\odot \left(\frac{M}{M_\odot}\right)^{0.27} \left(\frac{\dot M}{10^{-3} M_\odot \text{yr}^{-1}}\right)^{0.41}}
\label{eq:radius}
\end{equation}
where we smoothed $\rm{\dot M}$ by taking the average over the time taken to accrete 0.1 M$_\odot$. 
\ \

\subsection{Chemical network simulations}
\label{sec:chem}
\noindent The simulations originally presented \citetalias{Prole2023} were performed with the moving mesh code {\sc Arepo} \citep{Springel2010} with a primordial chemistry set-up. {\sc Arepo} solves hyperbolic conservation laws of ideal hydrodynamics with a finite volume approach, based on a second-order unsplit Godunov scheme with an exact Riemann solver. 
We used the same chemistry and cooling as \cite{Wollenberg2019}, which is described in the appendix of \cite{Clark2011}, but with updated rate coefficients, as summarised in \cite{Schauer2017}. The network has 45 chemical reactions to model primordial gas made up of 12 species: H, H$^{+}$, H$^{-}$, H$^{+}_{2}$ , H$_{2}$, He, He$^{+}$, He$^{++}$, D, D$^{+}$, HD and free electrons.  Included in the network are: H$_{2}$ cooling (including an approximate treatment of the effects of opacity), collisionally-induced H$_{2}$ emission, HD cooling, ionisation and recombination, heating and cooling from changes in the chemical make-up of the gas and from shocks, compression and expansion of the gas, three-body H$_{2}$ formation and heating from accretion luminosity. For reasons of computational efficiency, the network switches off tracking of deuterium chemistry\footnote{Note that HD cooling continues to be included in the model.} at densities above 10$^{-16}$~g~cm$^{-3}$, instead assuming that the ratio of HD to H$_{2}$ at these densities is given by the cosmological D to H ratio of 2.6$\times$10$^{-5}$. The adiabatic index of the gas is computed as a function of chemical composition and temperature with the {\sc Arepo} HLLD Riemann solver.

\begin{figure*}[!t]
	 \hbox{\hspace{-1.7cm} \includegraphics[scale=0.75]{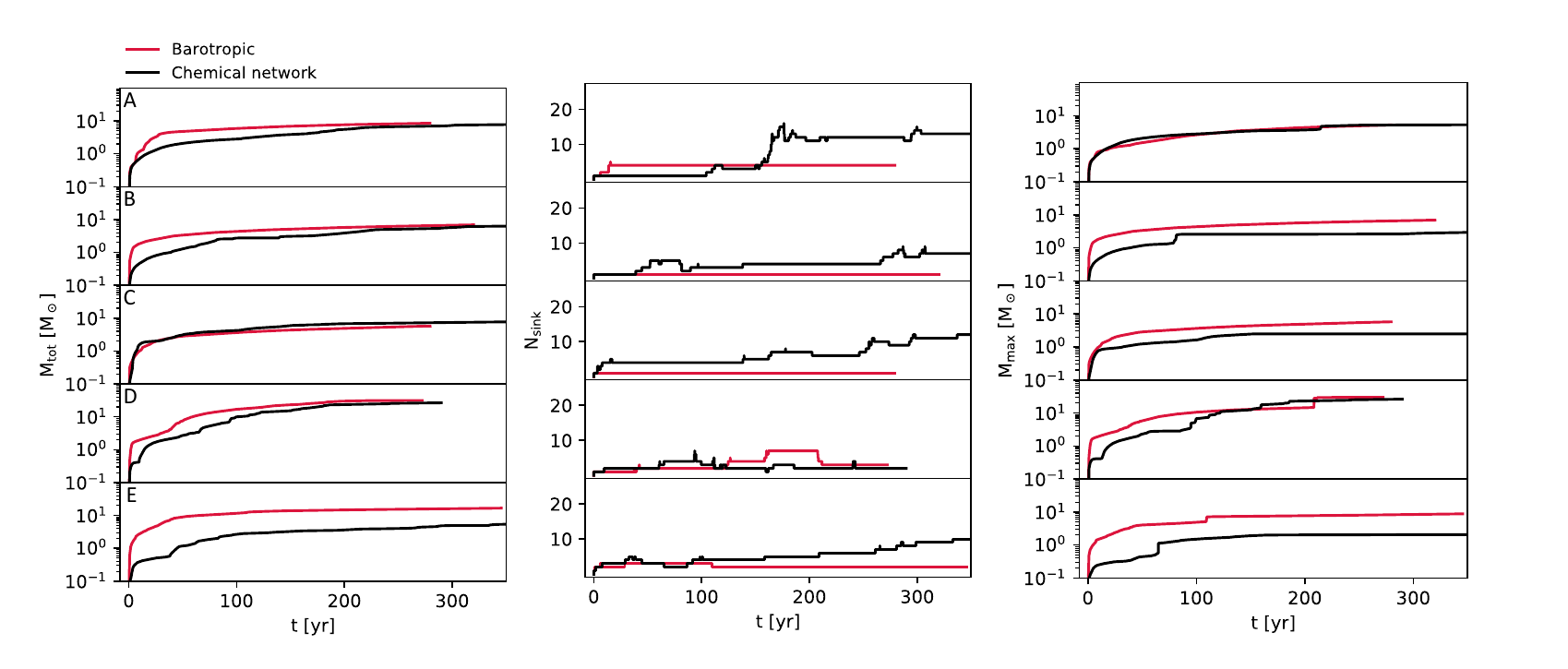}}
    \ \
    \caption{Comparison of CN runs with the barotropic EoS runs. From left to right -  total mass accreted across all sink particles, number of sink particles formed and mass of the most massive sink in all five halos as functions of time since the formation of the first sink particle. The time in each case refers to the time after the formation of the first sink in each realisation. In general the number of sinks formed when using an EoS is reduced, with the mass of the maximum sink when using the EoS higher  than in the CN case. The total mass in sinks is comparable between the CN and EoS realisations.}
\label{fig:frag}
\end{figure*}

\subsection{Barotropic EoS}
\label{sec:baro}
\noindent In this section we construct an EoS empirically from the simulations of \citetalias{Prole2023},  to replace the complex chemical network and reduce the computing time. 

To create a barotropic EoS for primordial star formation, we start by computing a mass-weighted temperature-density profile for each of the 5 halos from \citetalias{Prole2023} at a time just before the formation of the first sink particle, using a set of 50 logarithmically-spaced density bins. We refer to each halo with a letter from Halo A to E. We show these as blue lines in the left panel of Figure \ref{fig:eos}. We then take the mean temperature in each density bin across the 5 simulations to produce an average EoS, as shown by the black dashed line. To provide a temperature (and hence internal energy) to the gas cells given their density, we extrapolate linearly in log space between each of the 50 data points as

\begin{equation}
\rm log(T) = m_ilog(\rho)+c_i,
\end{equation}
where m and c are the gradient and offset, given by 
\begin{equation}
\rm m_i =  \frac{log(T_{i+1})-log(T_{i})}{log(\rho_{i+1})-log(\rho_{i})},
\end{equation}
and 
\begin{equation}
\rm c_{i}=log(T_{i}) - m_ilog(\rho_i)
\end{equation}
For densities lower than the first density bin $\rho_0$, we make the EoS flat at T$_0$. For densities higher than the last density bin, we continue along the same trajectory as log(T) = m$_{50}$log($\rho$)+c$_{50}$. 

The right panel of Figure \ref{fig:eos} shows the EoS contrasted against the temperature distribution produced by the CN for Halo E, for illustration purposes. The relation seemingly deviates from the distribution of temperatures at low densities because the 5 halos vary in mass and hence display different behaviours while coupled with the DM in this regime. However, the timescales involved with gas at those densities is much longer than the timescales for which the simulations can currently be run and hence can have no negative effect.

At densities above $\sim 10^{-13}$ g cm$^{-3}$, the range of temperatures at a given density becomes more constrained compared to lower densities. This dense gas is also the most computationally expensive to simulate, due to the high sound speed of the gas and due to the fact that load balancing of work becomes more problematic in regions of high cell count.

\begin{figure}[!b]
	 \hbox{\hspace{-1cm} \includegraphics[scale=0.7]{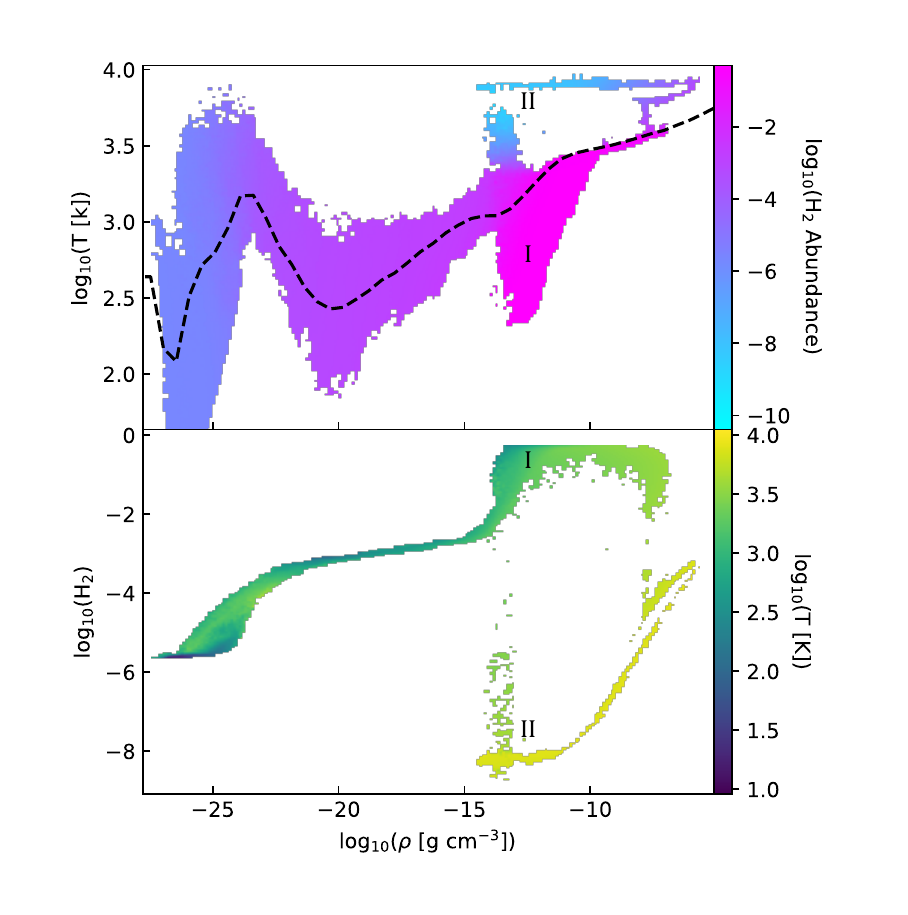}}
    \caption{The thermal state of the gas approximately 100 years after the formation of the first sink. Top Panel:  Temperature-density profile of the gas in the CN runs $\sim$ 100 yr after the formation of the first sink particle, coloured by H$_2$ mass fraction. The barotropic EoS is overplotted as a dashed black line. Regions marked I and II illustrate regions where the EoS fails to capture important thermodynamical processes. Bottom Panel: H$_2$ abundance as a function of density, coloured by temperature. The same regions are again marked and discussed in detail in the text.}
    \label{fig:off}
\end{figure}

\begin{figure*}
	 \hbox{\hspace{1.8cm} \includegraphics[scale=0.7]{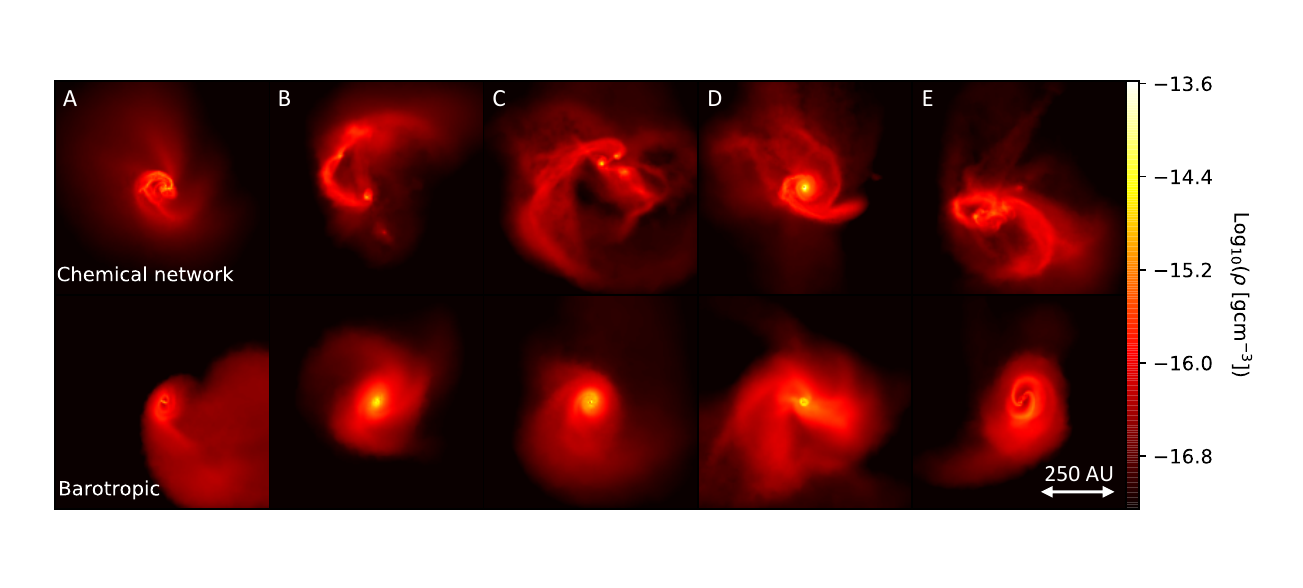}}
    \caption{Comparison of the disc structure in the chemical network (top) versus barotropic EoS runs (bottom). Column-weighted density projections of the inner 650 AU of the halos at a time approximately 300 yr after the formation of the first sink particle. The lack of sub-structure in the EoS runs is particularly noticeable.}
    \label{fig:proj}
\end{figure*}

\hfill \break
\section{Results}
\label{sec:results}

\noindent Halos A-E were re-simulated, starting from the last snapshot before the formation of the first sink particle ($\sim$4 yr before its creation) using our newly constructed EoS. For both the original CN and new EoS runs the collapse was followed up to a density of 10$^{-6}$ g cm$^{-3}$ before inserting sink particles and simulating a further 250 - 350 yr of fragmentation and accretion.
\subsection{Sink Particle Formation and Evolution}
Figure \ref{fig:frag} shows the total mass accreted across all sinks (left panel), the number of sinks formed (middle panel) and the mass of the most massive sink as a function of time since the formation of the first sink particle (right panel). Each simulation is evolved for between approximately 250 and 350 years after the formation of the first sink particle. In each panel the barotropic EoS is marked by a red line while the results when employing the full CN are marked as black lines. Difference between using the full CN and the EoS are immediately apparent.  Starting with the left panel we see that for the full CN model the total mass initially accreted, $\rm{M_{tot}}$, is lower in 4 out the 5 cases compared to the EoS model. However, over time the total mass accreted does converge between the EoS and CN models. This finding is consistent with what we see in the middle panel, where we show the cumulative number of sinks formed. Again for 4 out the 5 cases the total number of sinks formed is significantly higher in the CN model. In the other case, Halo D, the number of sinks formed in both models are comparable (and small). The EoS model consistently fails to form secondary protostars. Finally, in the right hand panel we show the mass of the most massive sink in each case. For all models the mass of the most massive sink in the EoS case exceeds that in the equivalent CN case. The reason being is that fragmentation is reduced in the EoS case allowing most gas to flow onto a smaller number of sinks. The formation of the first sink particle in the EoS runs is also delayed by periods of time ranging from a few to a few hundred years, as shown in Table \ref{table:1}, as a result of the changed chemical modelling. \\
\indent In summary, there is a significant reduction in gas fragmentation and subsequent formation of secondary sinks in the EoS compared to the CN model. The combined effect of these factors leads to the mass of the most massive sink in the EoS runs exceeding that in the corresponding CN run in 3 out of the 5 halos, as fewer sinks compete for the same amount of gas and fragmentation-induced starvation is avoided. In the other two models the mass of the final sink is comparable between the two models. The total mass in sinks is broadly similar in the EoS and CN cases i.e. the star formation rate is similar over the simulation time.

\begin{table}
	\centering
	\caption{Difference in formation time of the first sink particle between the chemical network and EoS runs, $\Delta$t = (t$_{\rm baro}$ - t$_{\rm CN}$), for each halo.}
	\label{table:1}
	\begin{tabular}{lr} 
		\hline
		Halo & $\Delta$t$_{\rm sink}$ [yr]\\
            \hline
		\ \ A & 381 \ \ \ \ \\
		\ \ B & 318 \ \ \ \ \\
  		\ \ C & 17 \ \ \ \ \\
		\ \ D & 38 \ \ \ \ \\
     	\ \ E & 354 \ \ \ \ \\
    	\hline
	\end{tabular}
\end{table}

\subsection{Fragmentation}
\noindent The difference in fragmentation behaviour can be explained by Figure \ref{fig:off}, which shows the temperature and H$_2$ abundance as a function of density $\sim$ 100 yr after the formation of the first sink particle in the CN runs, shown here for Halo E as an example. There are two significant deviations from the relation shown previously in Figure \ref{fig:eos}. Firstly, cold gas ($\sim$300 K) at densities of $\sim$10$^{-13}$ g cm$^{-3}$ (denoted as \MakeUppercase{\romannumeral 1} in Figure \ref{fig:off}) appears after the formation of the first protostar and was discussed in \cite{Clark2011}. The gas that initially collapses to high densities undergoes shock heating and rapid H$_2$ dissociation. However, once a rotating disc structure forms around the first protostar, infalling gas experiences less compression and retains its H$_2$ as shown in the bottom panel of Figure \ref{fig:off}. Collisions in the disc cause a sharp increase in temperature and density, which increases the H$_2$ formation rate and allows the gas to cool. The cold gas becomes Jeans unstable and fragments. In contrast, the temperature of the secondary gas falling in through the disc in the barotropic EoS runs does not experience this process of cooling and expansion and hence does not go on to fragment. This is the reason for reduced number of sink particles in the barotropic EoS runs compared to the CN runs.

The second noteworthy deviation from the EoS is the hot (10$^4$ K) gas at high densities (denoted as \MakeUppercase{\romannumeral 2} in Figure \ref{fig:off}), which is discussed in \cite{Stacy2010}. The intense gravitational potential of the protostar pulls gas towards it with velocities sufficiently high to heat the gas to these temperatures. The gas cannot heat to above these temperatures as cooling from atomic hydrogen begins to dominate over the adiabatic and viscous heating. At these temperatures H$_2$ molecules are dissociated \citep{Yoshida2008} as seen by the sliver of low H$_2$ abundance (high temperature) in the bottom panel.

The qualitative effects of these thermal changes are displayed in the density projections of the inner 650 AU of the halos, shown $\sim 300$ yr  after the formation of the first sink particle in Figure \ref{fig:proj}. Each panel refers to a different halo, from Halo A on the left to Halo E on the right. Although the size of the disc is invariant to the thermal treatment, the lack of unstable cold gas at high densities in the barotropic EoS runs results in reduced sub-structure in the disc compared to the CN runs, producing smooth discs which experience less fragmentation.

\subsection{Sink Particle Mass Distribution}

\begin{figure}
	 \hbox{\hspace{-0.62cm} \includegraphics[scale=0.81]{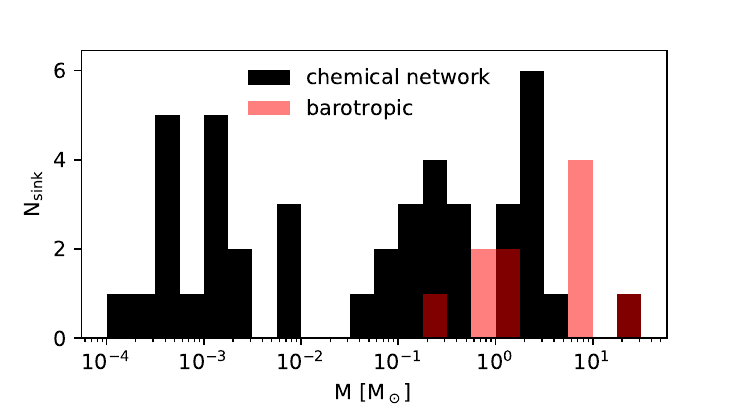}}
    \caption{The distibution of sink masses in the realisations run with the full CN compared to those run with the barotropic EoS. The sink masses from the CN runs are marked in black while those from the EoS are marked in red. The EoS realisations result in a much smaller distribution of masses and as well as much fewer sinks.}
    \label{fig:IMF}
\end{figure}

\noindent We show the combined distribution of sink particle masses across all 5 halos at a time $\sim$300 yr after the formation of the first sink in Figure \ref{fig:IMF}. The distribution summarises the results of this study; when using the EoS, the lack of cold gas collapsing through the disc after the formation of the first protostar leads to significantly reduced fragmentation and subsequent formation of secondary protostars, boosting the mass of the few protostars that do form through the lack of fragmentation-induced starvation. We note that the CN simulations produced a group of M$\rm{_{sink}}  < 0.075 \ $M$_\odot$ sink particles that were quickly ejected after their formation. These objects can be roughly interpreted as brown dwarfs but constitute only a small fraction of the mass in sinks. Metal-free and ultra metal-poor brown dwarfs have previously been reported in similar theoretical/numerical studies (e.g. \citealt{Machida2008e,Basu2012,Zhang2017c,Zhang2017b,Schlaufman2018}). These low mass protostars are not produced when using the EoS due to its inability to model disc fragmentation as already discussed.

\subsection{Computational Gains}
\begin{figure}
	 \hbox{\hspace{-0.85cm} \includegraphics[scale=0.77]{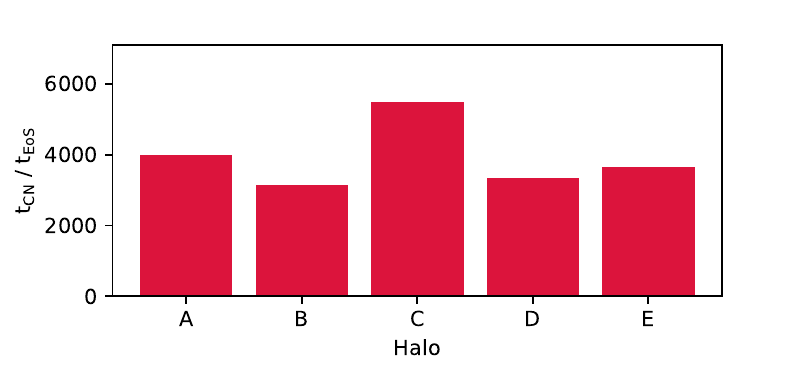}}
    \caption{Ratio of the wall-clock time taken to perform the CN and EoS routines. For each halo, we calculate the average time to perform the routine over 10 numerical timesteps after the formation of the first sink particle. On average the EoS model is 4000 times faster than solving the full CN.}
    \label{fig:speedup}
\end{figure}

\noindent Finally, we compare the physical time taken for the CN and EoS simulations. 
All simulations were performed on the supercomputer COSMA8\footnote{https://www.dur.ac.uk/icc/cosma/cosma8/}. The COSMA8 system consists of 360 compute nodes each with 1 TB RAM and dual 64-core AMD EPYC 7H12 water-cooled processors running at 2.6GHz. The large memory nodes make COSMA8 ideal for memory intensive calculations. The simulations here ran on 1 node using the full complement of 128 cores per simulation. \\
\indent Figure \ref{fig:speedup} shows the ratio of the average wall-clock times taken to perform the CN and EoS routines within the first 10 numerical timesteps after the formation of the first sink particle. The purpose of the barotropic EoS was to reduce the computational time, which it has by a factor of around 4000 due to the simplified calculation. This computational gain is not surprising given that the EoS calculation is extremely simple compared to solving the full network. As we discuss however in \S \ref{sec:summary} the EoS implemented here is somewhat simplistic and likely represents an upper limit to the possible speedups possible compared to a full chemistry network.


\ \

\ \ 
\section{Caveats}
\label{sec:caveats}
\noindent We have introduced our EoS at a time just before the formation of the first sink particle. At this point, all of the gas in the simulation box is forced onto the EoS. While the spread of temperatures around the EoS in the CN runs is tight for densities above 10$^{-13}$ g cm$^{-13}$, they cover a large range at lower densities. However, the dynamical times for gas at these densities is longer than the simulation time of a few hundred years and as such this should not effect the results.

We have not simulated the initial collapse of the gas into the halo using the EoS and hence our simulations do not indicate whether the EoS can accurately model the flow of gas into the halo and inital collapse phase. This study focused on the ability of the EoS to model the most computationally expensive part of the simulations which is the fragmentation of the gas after the formation of protostars. Studying the impact of using an EoS on the initial collapse is outside the scope of this work. Using an EoS, similar to that used here, may be possible for lower resolution simulations enabling longer modelling times. However, use of an EoS should be treated with caution as demonstrated here. 

We have resolved the gas using a refinement criteria of 16 cells per Jeans length. While purely hydrodynamic simulations typically require a lower refinement criteria than magneto-hydrodynamic simulations (e.g. \citealt{Schleicher2009,Turk2012}), it was recently noted that even purely hydrodynamic simulations require a refinement criteria of 64 cells per Jeans length to capture the thermal and chemical changes that occur across shocks where matter falls onto the disc (see Appendix of \citealt{Sharda2021}).

\section{Summary}
\label{sec:summary}
In this study we have explored whether the use of a barotropic EoS can produce a similar protostellar mass distribution to simulations using a full chemical network while reducing the computational resources required. To that aim, we have produced an empirically determined EoS by calculating the average temperature-density profile across five cosmological halos. We use the chemical network to simulate the initial H$_2$ production and cooling cycle as the gas collapses within the DM halo and the subsequent collapse to protostellar densities before switching to the EoS just before the formation of the first sink particle. The goal being to examine the impact that using an EoS has on the subsequent fragmentation, secondary protostar formation and overall star formation rate within Pop III minihalos.

The existence of multiple phases of gas at the same densities means that the fragmentation behaviour of the gas is not followed accurately by the EoS. The cold gas which collapses after the formation of the first protostar through its surrounding rotating disc is responsible for a large degree of fragmentation, which is not captured by the EoS formalism. The EoS method fails to capture the variation in the H$_2$ abundances which are crucial in tracking secondary fragmentation. We therefore conclude that the use of an EoS, in a similar way as done here, should be treated with caution.

We note that our EoS may be modified to more accurately predict the fragmentation behaviour by incorporating the different gas phases within the disc. In this case the secondary gas infalling through the already formed disc would be identified by its lack of compression (and shock-heating), allowing its temperature-density relationship to be changed to a alternative EoS more appropriate for the colder, Jeans unstable gas. While such a treatment is beyond the scope of this study, we hope to address future parameterisation accounting for the multiphase nature of the gas while retaining computational advantages in a future study (VandeBor et al. in prep). 

This result is relevant to ongoing Pop III studies which have commonly employed an EoS in place of a chemical network throughout the last two decades. Studies have modelled the collapse of metal-free gas using a simple polytropic EoS  (e.g. \citealt{Spaans2000,Saigo2004,Marassi2009,Riaz2018,Raghuvanshi2023}), an EoS fitted to a one-zone model (e.g. \citealt{Suda2007,Clark2008,Machida2008a,Machida2013,Machida2015,Susa2019}), or based on relativistic mean field theory (e.g. \citealt{Suwa2007a,Suwa2007}). All of these studies use EoSs similar to what we tested here. However, it should be noted that the 
regime in which the EoS is used will be a determining constraint but that careful estimation of possible limitations is necessary. 

We note that our method of switching to an EoS once the center of the halo reaches protostellar densities should not be confused with other studies which have employed a stiffened equation of state only to the gas at those high densities, effectively replacing sink particles by preventing the gas from attaining even higher densities, hence avoiding violation of the Truelove condition by preventing the gas' Jeans scale from shrinking passed the minimum cell scale (e.g. \citealt{Hirano2017,Hirano2022,Saad2022}). In these studies, the CN is still modelling the gas within the disc and hence the different phases shown in this study are included naturally.

In conclusion we find that using an empirically determined EoS can significantly speed up the computation but at the cost of not correctly modelling the multi-phase nature of the gas. We found that this resulted in significantly lower secondary protostar formation. Although the overall star formation rates and mass in sinks remained relatively unchanged. 

\section*{Acknowledgements}
We acknowledge support from the Irish Research Council Laureate programme under grant number IRCLA/2022/1165. JR also acknowledges support from the Royal Society and Science Foundation Ireland under grant number URF\textbackslash R1\textbackslash 191132. 

This work used the DiRAC@Durham facility managed by the Institute for Computational Cosmology on behalf of the STFC DiRAC HPC Facility (www.dirac.ac.uk). The equipment was funded by BEIS capital funding via STFC capital grants ST/P002293/1, ST/R002371/1 and ST/S002502/1, Durham University and STFC operations grant ST/R000832/1. DiRAC is part of the National e-Infrastructure.

We also acknowledge the support of the Supercomputing Wales project, which is part-funded by the European Regional Development Fund (ERDF) via Welsh Government.

Finally, we acknowledge Advanced Research Computing at Cardiff (ARCCA) for providing resources for the project.

\newpage
\bibliographystyle{mnras}
\bibliography{references_EoS}

\end{document}